\RequirePackage{lineno}
\documentclass[aps,prl,preprint,superscriptaddress]{revtex4}
\usepackage{graphicx}
\usepackage{amsmath}
\usepackage{color}

\begin{document}


\title{Two-photon pumped lead halide perovskite nanowire lasers}

\author{Zhiyuan Gu,$^\ddag$ Kaiyang Wang,$^\ddag$ Wenzhao Sun, Jinakai Li, Shuai Liu, Qinghai Song}
\email[]{qinghai.song@hitsz.edu.cn}
\affiliation{Department of Electronic and Information Engineering, Harbin Institute of Technology, Shenzhen, 518055, China}
\author{Shumin Xiao}
\email[]{shuminxiao@gmail.com}
\affiliation{Department of Material Science and Engineering, Harbin Institute of Technology, Shenzhen, 518055, China}

\date{\today}
\begin{abstract}
Solution-processed lead halide perovskites have shown very bright future in both solar cells and microlasers. Very recently, the nonlinearity of perovskites started to attract considerable research attention. Second harmonic generation and two-photon absorption have been successfully demonstrated. However, the nonlinearity based perovskite devices such as micro- \& nano- lasers are still absent. Here we demonstrate the two-photon pumped nanolasers from perovskite nanowires. The $CH_3NH_3PbBr_3$ perovskite nanowires were synthesized with one-step solution self-assembly method and dispersed on glass substrate. Under the optical excitation at 800 nm, two-photon pumped lasing actions with periodic peaks have been successfully observed at around 546 nm. The obtained quality (Q) factors of two-photon pumped nanolasers are around 960, and the corresponding thresholds are about $674 \mu J/cm^2$. Both the Q factors and thresholds are comparable to conventional whispering gallery modes in two-dimensional polygon microplates. Our researches are the first demonstrations of two-photon pumped nanolasers in perovskite nanowires. We believe our finding will significantly expand the application of perovskite in low-cost nonlinear optical devices such as optical limiting, optical switch, and biomedical imaging et al.
\\
{\bf $^\ddag$ these authors contribute equally to this research}

\end{abstract}

\maketitle
\section{Introduction}

In past three years, solution processed perovskites have attracted intensively research attention~\cite{xing,xing2,Xing3,yang,Ha,Zhu,Zhang,Liao,shi,wehren,dong,liux}. Due to their long carrier lifetimes and diffusion lengths, perovskites have shown great potentials in low-cost and high efficient solar cells. The light conversion efficiency has been quickly improved and the record value reaches $20.1\%$ very recently~\cite{yang}. Besides the efficient light harvesting, perovskites have also been revealed as promising gain materials in 2014~\cite{xing2,Ha}. Because the gain of perovskites are strongly dependent on the crystal quality and the structure of materials~\cite{Zhu,Xing3}, perovskites with reduced dimensions have been quickly developed and lasing actions with low thresholds have been successfully demonstrated in two-dimensional square~\cite{Liao} and hexagonal~\cite{Zhang} microdisks. In 2015, single crystal one-dimensional nanowires have been synthesized to further improve the characteristics of nanolasers ~\cite{Zhu,Xing3}. Record low threshold ($\sim 220 nJ/cm^2$) and high quality (Q) factor were realized in single lead halide perovskite nanowire~\cite{Zhu}. These perovskite nanowires are expected to play important role as building blocks for nanoscale photonic and optoelectronic devices~\cite{yan}. They have also enabled the possibility of bridging future perovskite nano-networks with conventional systems.

More than the superior properties in solar cell and light emission, perovskites also have great potentials in low-cost, solution processed nonlinear optical devices, which provide a new opportunity for both perovskites and nonlinear optics~\cite{Stou,Walters}. Stoumpos et al. have studied the second harmonic generation (SHG) of organic/inorganic germanium perovskite~\cite{Stou}. Very large second order nonlinear susceptibility and high damage threshold have been demonstrated. Walters et al. reported the two-photon absorption of organometallic bromide perovskites and the obtained two-photon absorption efficiency is even comparable to CdSe and CdS~\cite{Walters}. These progresses have triggered great research attentions on nonlinear perovskite devices. In particular, there has been great interest in two-photon absorption effects of nanomaterials due to their applications in imaging, nanolaser, and optical limits. For example, the short penetration depth of single-photon pumping laser can be resolved by two-photon absorption~\cite{Sun} and thus makes perovskites to be potentially used in complex environment such as biological tissues. Here, as a first step, we demonstrate the multiphoton pumped lasing actions in lead halide perovskite nanowires. Lasing actions have been observed in perovskite nanowires that were optically excited via two-photon excitation. The Q factors and thresholds of two-photon pumped nanolasers are even comparable to the conventional whispering gallery modes in polygon shaped microdisks.

\section{Experimental results}
\subsection{Synthesis of perovskite nanowire}

Our lead halide perovskites were synthesized with one-step solution self-assembly method, which has been reported by Liao et al recently~\cite{Liao,shi}. Basically, $CH_3NH_3Br$ and $PbBr_2$ were independently solved in N,N-dimethylformamide (DMF) with concentrations around 0.1 M. Then two solutions were mixed at room temperature with 1:1 volume ratio to form $CH_3NH_3Br \cdot PbBr_2$ solution (0.05 M). The diluted solution was dip-casted onto a glass substrate, which was placed on a teflon stage in beaker. 15 ml dichloromethane (DCM) of $CH_2Cl_2$ was placed in the beaker and sealed with a porous Parafilm (3 M) to control the evaporation speed. After 24 hours, lead halide perovskites ($CH_3NH_3PbBr_3$) have been successfully synthesized on the substrate. By optically exciting with Ti:Sapphire laser at 800 nm (regenerated, Spectra physics, seeded by Maitai), the synthesized perovskites have been measured with fluorescent microscope. As shown in Fig. 1(a), the perovskites are dominated by two types of structures. One is the rectangle shaped microplates. The length of rectangle plate changes from a few micron to tens of micron. The other one is perovskite nanowire. By characterizing the nanowire with scanning electron microscope (SEM), we know that their lengths and widths are within the ranges of 10 - 30 microns and $500 nm$ - $1500 nm$, respectively (see the supplementary information). Following the recent reports of Zhu~\cite{Zhu} and Xing~\cite{Xing3}, these perovskite nanowires are large enough to support the nanolasers.

\subsection{Lasing actions in perovskite nanowires}

\begin{figure}[htb]
\includegraphics[width=0.8\textwidth]{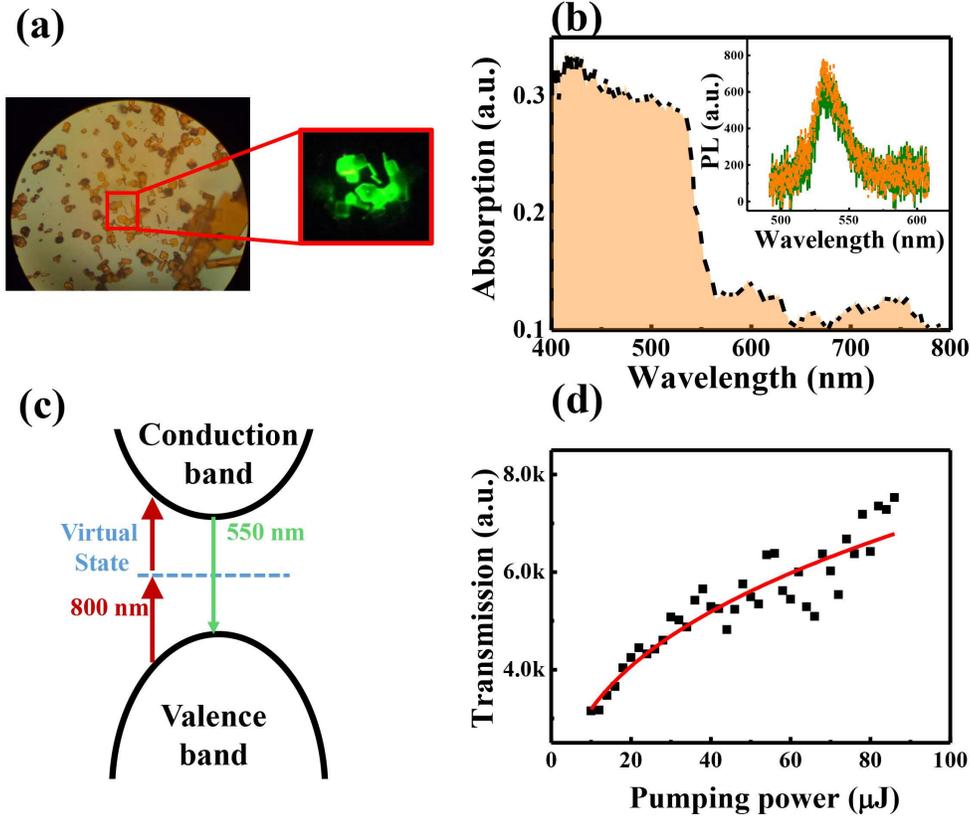}

\caption{Optical properties of perovskites. (a) Microscope image and florescent microscope images of perovskites under two-photon excitation at $600 \mu J/cm^2$. (b) Linear absorption spectra (black dashed line). The inset in (b) shows the photoluminescence spectra of perovskites under single-photon excitation (green solid line) and two-photon excitation (orange dash dotted line). Here the pumping densities of single photon and two-photon pumping are  $2.75 \mu J/cm^2$ and $600 \mu J/cm^2$, respectively. (c) Schematic picture of the two-photon absorption. (d) The transmission of ultrashort pulse at 800 nm as a function of incident power. Two-photon induced absorption can be clearly observed at large pumping power.}
\end{figure}

The linear properties of perovskites has been studied at the first. The dashed line Fig. 1(b) shows the recorded spectrum of linear absorption of a single perovskite microdisk. We can see that the absorption is very low at longer wavelength. The background and fluctuation at longer wavelength are induced by the reflection of perovskite and scattering of micron-sized perovskite objects. Once the wavelength is smaller than 540nm, the absorption increases quickly and reaches a near constant level. Then it is easy to know that the bandgap of the synthesized perovskite is around 2.29 eV. This value is consistent with the recent repots about perovskite crystals well~\cite{Walters,Liao}. From the absorption spectrum, we also know that the pumping laser at 800 nm is far below the bandgap of perovskite (see the schematic picture of two-photon absorption in Fig. 1(c)). Meanwhile, as the ultrashort laser pulse with low repetition rate (100fs, 1kHz) has been applied here, the photoinduced absorption associated with excited-state carriers can be neglected~\cite{shi}. In this sense, perovskites shall be transparent to the incident laser under the linear model. Figure 1(d) shows the dependence of transmission of Ti:Sapphire laser at 800 nm on the incident power. Different from the linear model, we can see that the output is not linearly dependent on the incident power. A clear optical-limiting behavior can be observed at high pumping power, indicating the nonlinear absorption in perovskites very well. Meanwhile, green light can also be observed under the florescent microscope (see Fig. 1(a)as an example), demonstrating the two-photon pumped photoluminescence well.

In the lasing experiment, one perovskite nanowire has been selected to study the detail laser behaviors. The structural information of the nanowire are shown in Figs. 2(a) and 2(b). The width and length of nanowire are 840 nm and 18.64 $\mu m$, respectively. And the thickness is around 625 nm. The nanowire was placed in a home-made microscope and first excited with frequency doubled Ti:Sapphire laser (at 400 nm) to determine the resonant properties and corresponding lasing actions (see supplementary information). In our experiment, the pumping laser was focused by a 40X objective lens onto the top surface of nanowire~\cite{Wangky}. The emitted lights were collected by the same objective lens and coupled to a CCD (Princeton instrument, PIXIS BUV) coupled spectrometer (Acton SpectroPro2700i) via a multimode fiber. The green line in Fig. 1(b) is the recorded spectrum of photoluminescence, which is a broad peak centered at 534 nm with a full width half maximum (FWHM) around 20nm. When the pumping density was increased, the emission intensity also increased and sharp peaks appeared in the spectrum. One example is shown in Fig. 2(c) with pumping density at 3.65 $\mu J/cm^2$. More than 10 periodic laser peaks can be observed within spectral range around 546 nm. Here the lasing wavelength happened at the right side of photoluminescence peak, which was caused by the rapidly increased absorption of perovskite (see Fig. 1(b)) at shorter wavelength. The mode spacing is around 1.486 nm and the FWHMs of sharp peaks are found to be 0.604 nm. The inset in Fig. 2(c) shows the fluorescence microscope image of perovskite nanowire. Bright spots can be observed at two end-facets of nanowire. The formation of bright spots well indicates the occurrence of amplification within the nanowire (the amplification length is the longest along the nanowire)~\cite{Zhu}.

\begin{figure}[htb]
\includegraphics[width=0.8\textwidth]{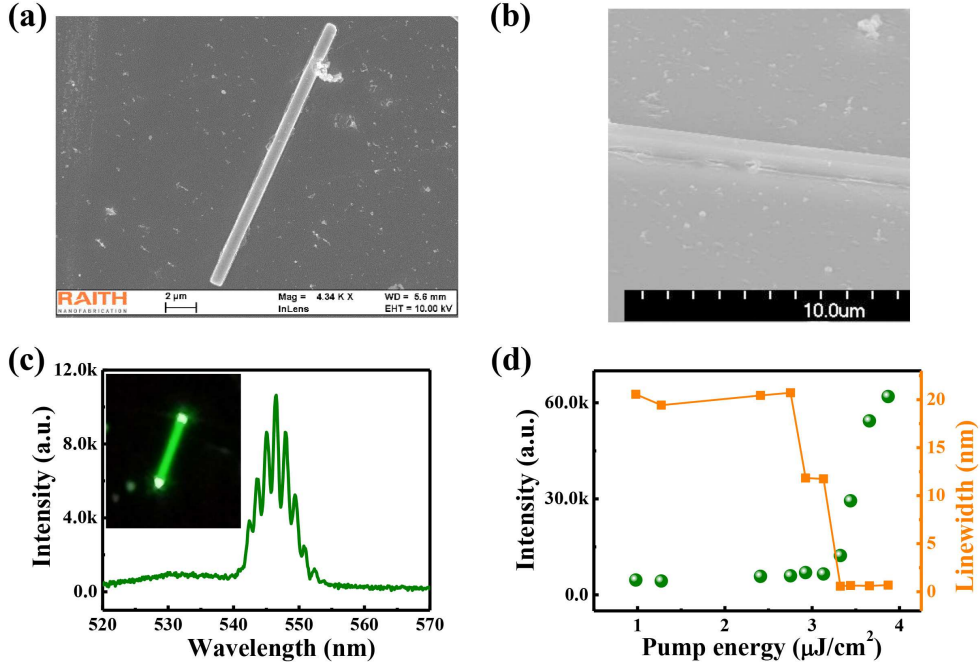}
\caption{Lasing actions in perovskite nanowire under single-photon excitation. (a) top-view and (b) tilt-view SEM images of perovskite nanowire. (c) Emission spectrum from perovskite nanowire. Inset is the corresponding fluorescence microscope image. Here the pumping density is $3.65 \mu J/cm^2$. (d) The output intensity and linewidth as a function of pumping density.}
\end{figure}

To determine the onset of lasing actions, we have studied the dependence of output intensities on the pumping density. All the results are shown in Fig. 2(d). When the pumping density was increased from 0.98 $\mu J/cm^2$ to 3.14 $\mu J/cm^2$, the intensity increased slowly. Once the pumping intensity was further increased, the emission intensity was dramatically enhanced. And the FWHM of emission spectra reduced simultaneously, indicating a clear laser threshold at around 3.14 $\mu J/cm^2$. Therefore, lasing actions within one-step solution-processed perovskite nanowire can be confirmed. We note that the threshold in Fig. 2(d) is about one order of magnitude higher than the smallest value for $CH_3NH_3PbBr_3$ perovskite nanowire in literature ($\sim$ 300 $nJ/cm^2$)~\cite{Zhu}. It is already lower than the thresholds of square shaped microdisks that are synthesized in the same process~\cite{Liao}. These differences in different examples and experiments might be caused by the crystal quality and cavity size~\cite{Zhu,Xing3}.

According to Walters et al.'s recent study, the two-photon absorption coefficient ($\sim$ 8.6 $cm GW^{-1}$) of single-crystal perovskites is similar to the conventional inorganic semiconductor CdS and CdSe~\cite{Walters}. Thus perovskites can be an excellent candidate for two-photon-excited up-conversion devices, especially for the nanowire lasers that have ultralow thresholds~\cite{Zhu,Xing3}. We then removed the BBO crystals and directly pumped the same nanowire with 800nm Ti:Sapphire laser. Interestingly, green lights can be clearly observed under fluorescent microscope (see Fig. 1(a) as an example). The corresponding spectrum of photoluminescence is depicted as orange line in Fig. 1(b). A broad emission peak with FWHM around 20 nm has also been observed under fluorescent microscope. In this nanowire, the emission spectra under single-photon and two-photon excitations are very close. In some cases, the two-photon pumped spectrum slightly shifted to longer wavelength (see supplementary information)~\cite{Yu}.

\begin{figure}[htb]
\includegraphics[width=0.8\textwidth]{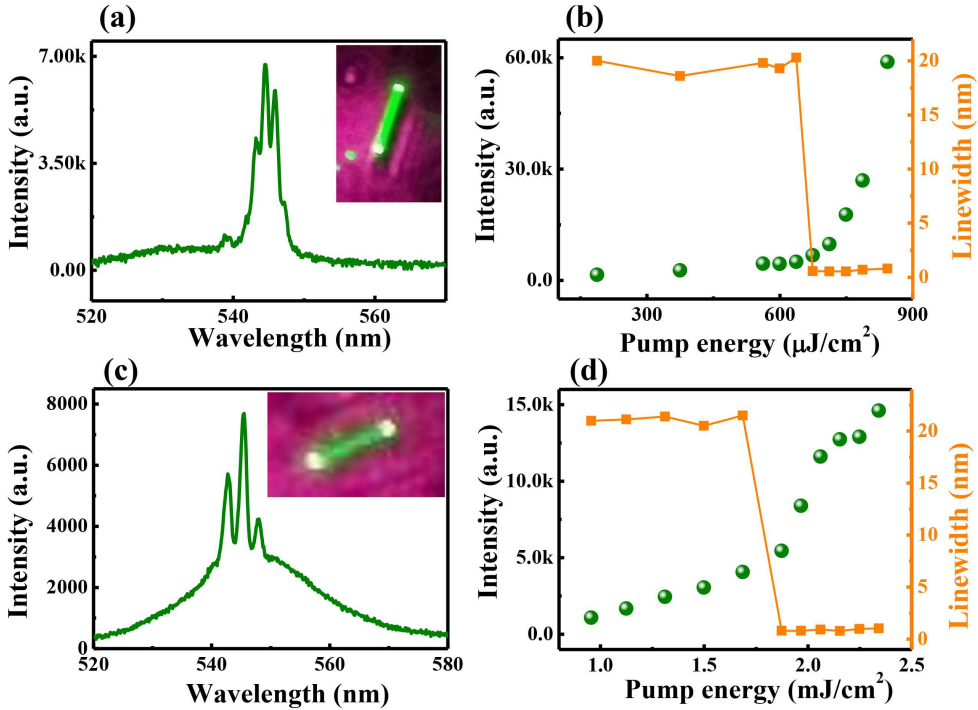}

\caption{Lasing actions in perovskite nanowire under two-photon excitation. (a) Emission from perovskite nanowire under two-photon excitation at $786 \mu J/cm^2$. Inset shows the fluorescent microscope image. (b) The dependence of output intensity and linewidth as a function of pumping density. (c) and (d) shows the lasing spectrum and threshold behaviors of a different nanowire. Inset in (c) also shows the corresponding fluorescent microscope image. Here the pumping density is 2.24 $mJ/cm^2$}
\end{figure}

With the increase of pumping power, some dramatic changes could be seen in fluorescent images. Different from the relative uniform fluorescence image (Fig. 1(a)), bright spots started to appear at two ends of nanowire (inset in Fig. 3(a)), indicating the onset of lasing actions~\cite{Zhu}. One of the emission spectrum was recorded and plotted in Fig. 3(a). Similar to the single-photon pumped spectrum, periodic lasing peaks can be observed. Here both the mode spacing ($\sim 1.49 nm$) and the lasing wavelengths are very close to the spectrum in Fig. 2(c). The only difference is that the spectral range of two-photon pumped lasers is narrower than the one under single-photon excitation~\cite{Yu,Hou}. Only three main lasing peaks can be observed in Fig. 3(a). This is also caused by the intrinsic absorption of perovskite at the lasing wavelength range~\cite{Yu}.

The laser threshold of two-photon pumped nanowire laser has also been studied. Figure 3(b) shows the output intensity as a function of pumping density. When the pumping density was smaller than 674 $\mu J/cm^2$, the output intensity increased slowly. As we mentioned above, this process corresponded to spontaneous emission (see spectrum in orange line in Fig. 1(b)). Once the pumping density was above 674 $\mu J/cm^2$, the output intensity increased dramatically. Meanwhile, sharp peaks emerged in laser spectra (see Fig. 3(b)) and the FWHM quickly reduced from around $\sim 20 nm$ to below 1 nm. Thus the lasing behaviors of two-photon pumped nanowire can be easily confirmed. Here the threshold value (674 $\mu J/cm^2$, comparable to the threshold of hexagonal microdisk lasers) of two-photon pumped nanowire laser is about 200 times of the one under single-photon excitation. This is also consistent with the observations of two-photon pumped lasing actions in CdS and CdSe. At the threshold point, where the gain and loss and supposed to be balanced, the linewidth is as small as 0.57 nm, corresponding to a radiation Q factor around 960. This high Q factor is even comparable to the recently reported lasers in perovskite squares and hexagon shaped microcavities. ¡¡
When the pumping power was above threshold, the laser linewidth increased a little bit and the laser peak slightly blueshifted simultaneously. As we mentioned before, this is induced by band filling effect during the ultrashort lasing pumping~\cite{Song}.

We note that two-photon pumped lasing actions are very generic in perovskite. It has been observed in almost all the nanolasers that produce laser emissions under single-photon excitation. One additional example is shown in Fig.s. 3(c) and 3(d) (There are the other 9 samples in the supplementary information). When a different nanowire was pumped under Ti:Sapphire laser at 800 nm, the emission spectrum consists of periodic lasing peaks and bright spots can be observed at the ends of nanowire (inset in Fig. 3(c)). Figure 3(d) shows that the emission intensity increased significantly and the FWHM reduced dramatically when the pumping power was above threshold ($1.8 mJ/cm^{2}$). All these phenomena are similar to Figs. 3(a) and 3(b) well. Slight differences happen in threshold value and FWHM at threshold point ($\sim 0.8 nm$, corresponding to $Q \sim 690$).These differences are caused by the relatively short length of the second perovskite nanowire.

\section{Discussions}
\begin{figure}[htb]
\includegraphics[width=0.8\textwidth]{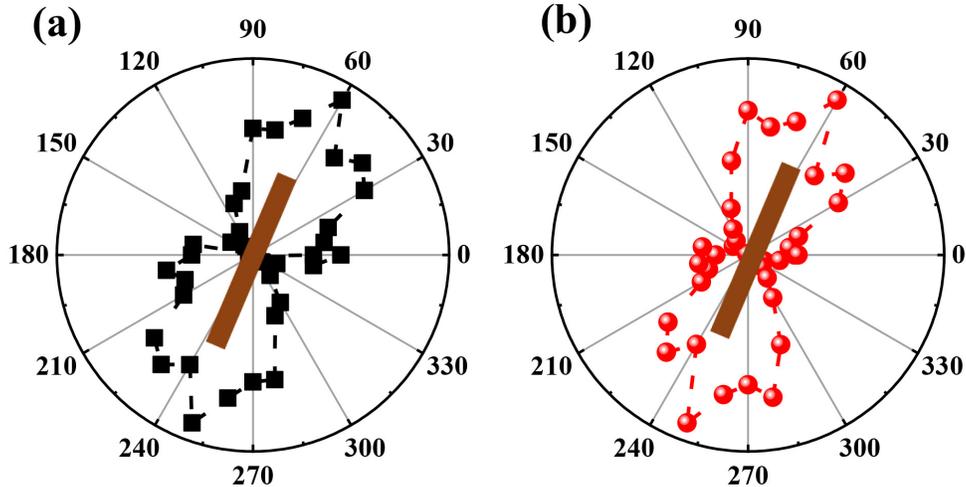}

\caption{Polarization and mode spacing of odd (a) and even (b) order perovskite nanowire lasing modes.}
\end{figure}

After the observations of lasing actions in perovskite nanowire with single-photon and two-photon pumping, it is also interesting to understand the mechanism that is responsible for the periodic lasing modes in Figs. 2 and 3. In general, the periodic lasing peaks in perovskite nanowire are generated from the Fabry-Perot resonances along the fundamental waveguide mode~\cite{Zhu}. This is also consistent with the bright spots at the ends of nanowires (see the insets in Figs. 2 and 3). The mode spacing of Fabry-Perot cavity can be easily estimated from the equation ($\Delta\lambda = \lambda^2/2nL$, where $n$ is the refractive index of perovskite and $L$ is the length of nanowire). For the nanowire in Fig. 2(a), the estimated value is $\Delta\lambda = 3.18$ nm, which is almost twice of the experimentally measured one. Such kind of discrepancy can be explained by taking account of two sets of Fabry-Perot modes along different waveguide modes. However, it is still not clear which modes are involved in the lasing actions. This is because that the width and thickness of nanowire are several times of lasing wavelength ($\lambda / n \sim 217 nm$). The perovskite nanowire can support more than fifty waveguide modes (see examples in the supplementary information).

To build a correct model for our nanolaser, we thus experimentally studied the polarizations of the perovskite lasers. Since the measured mode spacing is about half of the estimated value, we separated the odd and even number lasing modes into two groups and studied their polarizations independently. All the results are shown in Fig. 4. We can see that two types of resonances are both polarized along the nanowire, clearly demonstrating the transverse magnetic (TM, E is perpendicular to the substrate) polarizations. Thus in the numerical model, we can only consider the TM polarized resonances. In addition, in the SEM images in Figs. 2(a) and 2(b), there are a lot of nanoparticles that are randomly distributed around the nanowire. Many of them are even attached onto the surfaces of perovskite nanowire. Consequently, strong scattering at the surfaces can be expected. Since higher order waveguide modes usually have stronger field distributions at the surfaces, these modes usually experience larger scattering loss and can be suppressed in lasing experiments.

Following above analysis, we then numerically studied the eigenfrequencies of perovskite nanowire by full three-dimensional calculations with finite difference time domain (FDTD) method ~\cite{Guzy} and finite element method (COMSOL Multiphysics 4.3a). In our calculation, we mainly studied the TM polarized Fabry-Perot modes in the nanowire. Here the refractive indices of nanowire and substrate are fixed at $n = 2.55$~\cite{Loper} and $n = 1.5$, respectively. Figure 5(a) shows the numerically calculated transmission spectrum along the fundamental waveguide mode ($TM_{00}$, see the field pattern in the inset) in transverse plane of the perovskite nanowire. Within the spectral range from 540 nm to 560 nm, we can see 7 discrete peaks. The average mode spacing is about 3.1 nm, which is similar to the estimation of Fabry-Perot modes. Figure 5(b) shows the transmission spectrum along first order TM waveguide mode ($TM_{01}$, see see the field pattern in transverse plane). Similar to Fig. 5(a), another 6 discrete peaks can also be observed. Interestingly, these peaks appear around the center of every two peaks in Fig. 5(a). Consequently, if we consider these two types of Fabry-Perot modes at the same times, the average mode spacing is thus decreased to $\sim 1.55 nm$, which is very close to the experimental results ($\sim 1.49 nm$) and verifies our assumed mechanism well. The slight difference between experiment (2.98 nm) and numerical simulation (3.1 nm) can be explained by the group index of perovskite~\cite{Lebental}.

\begin{figure}[htb]
\includegraphics[width=0.8\textwidth]{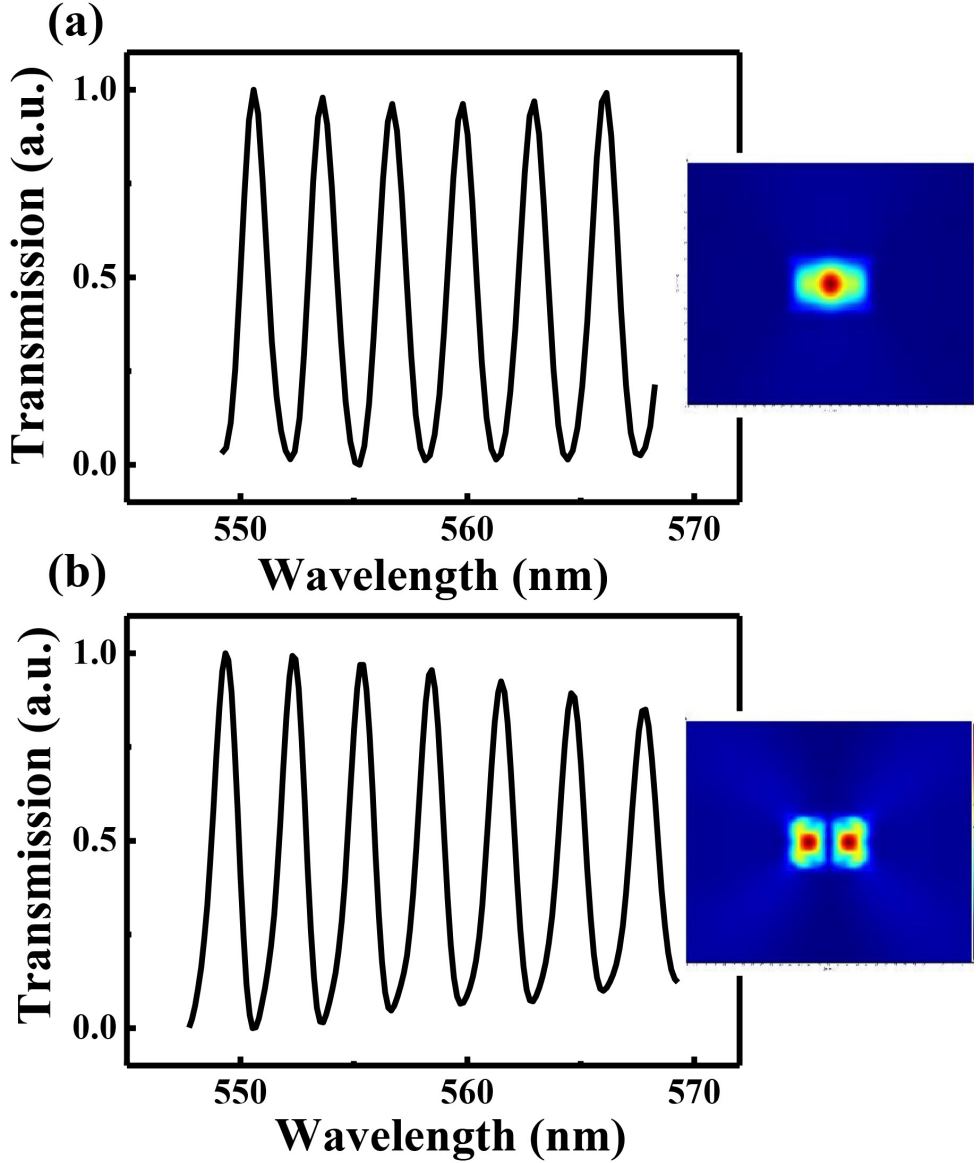}

\caption{Numerical simulation of resonances in perovskite nanowires. (a) Transmission spectrum of Fabry-Perot modes along the fundamental waveguide mode. Inset shows the field distribution of mode-1 in cross sections of x-y plane. (b) Transmission spectrum of Fabry-Perot modes along first order waveguide mode. The field profiles in transverse plane is shown in the inset.}
\end{figure}

Since there are too many waveguide modes in the transverse plane, one may argue that the lasing modes might come from two sets of higher order waveguide modes. This possibility can be excluded by the mode spacing well. While more than fifteen waveguide modes exist, their effective refractive indices are quite different. In general, the effective refractive indices of higher order waveguide modes reduce quickly from $\sim 2.5$ ($TM_{00}$) to $\sim 2$ ($TM_{04}$ or $TM_{22}$) (see supplementary information). Therefore, the Fabry-Perot modes along these waveguide modes shall have much larger mode spacings than the ones in Fig. 5. Then the difference between mode spacing in numerical simulation and experiment will be more dramatic and hard to be explained.


Another interesting phenomenon lies in the Q factors. The experimentally measured Q factor is about 960. The numerically calculated Q ($Q \sim \lambda/\Delta\lambda$) factors in Fig. 5(a) are also around 400. Then the numerical calculations and experimental results match well. And this is the fundamental basis of the relative low thresholds of perovskite nanowire lasers. However, both the experimentally measured Q factors and the numerically calculated Q factors are much larger than estimated values following the equation of Fabry-Perot modes ($Q = -(Lk_a)/\ln|r^2| \sim 100$)~\cite{Guzy2}. This is because that the resonant modes are not normally incident onto the end-facets as above consideration. From their propagating constants, we can estimate the incident angles on the endfacets as $11.74^{o}$ and $16.63^{o}$, respectively. By taking account the incident angles, the reflections at two end-facets are enhanced and thus the corresponding Q factors are improved to 411 and 526, which match the numerical calculation and experimental results well. Then the relative low thresholds of both single-photon and two-photon pumped perovskite nanowire lasers are also understandable.

\section{conclusion}
In conclusion, we have studied the lasing actions in perovskite nanowire. When the nanowire was optically excited by lasing beam at 400 nm (single-photon) and 800nm (two-photon), laser emissions have both been observed. Both the recorded Q factors and thresholds of two-photon pumped nanowire laser are even comparable to some single-photon excited polygon microlasers. This is the first time that two-photon pumped lasers in perovskites have been experimentally observed. Moreover, our experimental results are not limited in particular perovskite nanowire. Similar phenomena have also been observed in a number of different perovskite nanowires (see one example in supplementary information) and perovskite rectangle microdisks. Our experimental results clearly confirmed the third order nonlinearity of perovskites and made perovskite nonlinear optics devices to be promising. Considering the long penetration depth of laser at $600 - 1300 nm$, our observations shall also have potential applications in biomedical imaging and pinpoint detection.

\begin{acknowledgments}

This work is supported by NSFC11204055, NSFC61222507, NSFC11374078, NCET-11-0809, KQCX2012080709143322, KQCX20130627094615410, JCYJ20140417172417110, and JCYJ20140417172417096.

\end{acknowledgments}


\begin{thebibliography}{99}
\bibitem{xing} G. Xing, N. Mathews, S. Sun, S. S. Lim, Y. M. Lam, M. Gr\"{a}tzel, S. Mhaisalkar, and T. C. Sum, {\it Science} {\bf 2013}, {\it 342}, 344.
\bibitem{yang} W. S. Yang, J. H. Noh, N. J. Jeon, Y. C. Kim, S. Ryu, J. Seo, and S. I. Seok, {\it Science} {\bf 2015}, {\it 348}, 1234-1237.
\bibitem{xing2} G. Xing, N. Mathews, S. S. Lim, N. Yantara, X. Liu, D. Sabba, M. Gr\"{a}tzel, S. Mhaisalkar, and T. C. Sum, {\it Nat. Mater.} {\bf 2014}, {\it 13}, 476.
\bibitem{Ha} S. T. Ha, X. Liu, Q. Zhang, D. Giovanni, T. C. Sum, and Q. Xiong, {\it Adv. Opt. Mater.} {\bf 2014}, {\it 2}, 838.
\bibitem{Zhu} H. Zhu, Y. Fu, F. Meng, X. Wu, Z. Gong, Q. Ding, M. V. Gustafsson, M. Tuan Trinh, S. Jin, and X-Y. Zhu, {\it Nat. Mater.} {\bf 2015}, {\it 14}, 636.
\bibitem{Xing3} J. Xing, X. F. Liu, Q. Zhang, S. T. Ha, W. W. Yuan, C. Shen, T. C. Sum, and Q. Xiong, {\it Nano Lett.} {\bf 2015}, {\it 15}, 4751.
\bibitem{Liao} Q. Liao, K. Hu, H. H. Zhang, X. D. Wang, J. N. Yao, and H. B Fu, {\it Adv. Mater.} {\bf 2015}, {\it 27}, 3405.
\bibitem{Zhang} Q. Zhang, S. T. Ha, X. Liu, T. C. Sum, Q. Xiong, {\it Nano Lett.} {\bf 2014}, {\it 14}, 5595.
\bibitem{shi} D. Shi, V. Adinolfi, R. Comin, M. J. Yuan, E. Alarousu, A. Buin, Y. Chen, S. Hoogland, A. Rothenberger, K. Katsiev, Y. Losovyj, X. Zhang, P. A. Dowben, O. F. Mohammed, E. H. Sargent, and O. M. Bakr, {\it Science} {\bf 2015}, {\it 347}, 519.
\bibitem{dong} Q. Dong, Y. Fang, Y. Shao, P. Mulligan, J. Qiu, L. Cao, and J. Huang, {\it Science} {\bf 2015}, {\it 347}, 967.
\bibitem{liux} X. Liu, S. T. Ha, Q. Zhang, M. de la Mata, C. Magen, J. Arviol, T. C. Sum, and Q. Xiong, {\it ACS Nano} {\bf 2015}, {\it 9}, 687-695.
\bibitem{wehren} C. Wehrenfennig, M. Liu, H. J. Snaith, M. B. Johnson, and L. M. Herz, {\it J. Phys. Chem. Lett.} {\bf 2014}, {\it 5}, 1300-1306.
\bibitem{yan} R. X. Yan, D. Gargas, and P. D. Yang, {\it Nat. Photon.} {\bf 2009}, {\it 3}, 569-576.
\bibitem{Stou} C. C. Stoumpos, L. Frazer, D. J. Clark, Y. S. Kim, S. H. Rhim, A. J. Freeman, J. B. Ketterson, J. I. Jang, and M. G. Kanatzidis, {\it J. Am. Chem. Soc.} {\bf 2015}, {\it 137}, 6804-6819.
\bibitem{Walters} G. Walters, B. R. Sutherland, S. Hoogland, D. Shi, R. Comin, D. P. Sellan, O. M. Bakr, and E. H. Sargent, {\it ACS Nano} {\bf 2015}, 10.1021/acsnano.5b03308.
\bibitem{Sun} C. Zhang, C. L. Zou, Y. L. Yan, R. Hao, F. W. Sun, Z. F. Han, Y. S. Zhao, and J. N. Yao, {\it J. Am. Chem. Soc.} {\bf 2011}, {\it 133}, 7276.
\bibitem{Wangky} K. Y. Wang et al., {\it ACS Photonics} {\bf 2015}, {\it 2}, 1278-1283.
\bibitem{Yu} J. Yu, Y. Cui, H. Xu, Y. Yang, Z. Wang, B. Chen, and G. Qian, {\it Nat. Comm.} {\bf 2013}, {\it 4}, 2719.
\bibitem{Hou} K. Hou, Q. H. Song, D. B. Nie, F. Y. Li, Z. Q. Bian, L. Y. Liu, L. Xu, C. H. Huang, {\it Chem. Mater.} {\bf 2008}, {\it 20}, 3814.
\bibitem{Song} Q. H. Song, H. Cao, S. T. Ho, G. S. Solomon, {\it Appl. Phys. Lett.} {\bf 2009}, {\it 94}, 061109.
\bibitem{Guzy2} Z. Y. Gu, N. Zhang, Q. Lyu, K. Y. Wang, S. M. Xiao, and Q. H. Song, {\it Laser and Photonics Reviews}, Arxiv.
\bibitem{Guzy} Z. Y. Gu, S. Liu, S. Sun, K. Y. Wang, Q. Lyu, S. M. Xiao, Q. H. Song, {\it Sci. Rep.} {\bf 2015}, {\it 5}, 9171.
\bibitem{Loper} P. Loper et al., {\it J. Phys. Chem. Lett.} {\bf 2015}, {\it 6}, 66-71.
\bibitem{Lebental} M. Lebental, N. Djellali, C. Arnaud, J. S. Lauret, J. Zyss, R. Dubertland, C. Schmit, and E. Bogomolny, {it Phys. Rev. A} {\bf 2007}, {\it 76}, 023830.

\end{thebibliography}
\end{document}